\documentstyle[a4,12pt,epsfig]{article}
\begin{document}

\newcommand \be  {\begin{equation}}
\newcommand \ee  {\end{equation}}

\title{\bf ARE FINANCIAL CRASHES PREDICTABLE?}

\author{Laurent Laloux$^1$, Marc Potters$^1$, Rama Cont$^{1,2}$,\\
Jean-Pierre Aguilar$^{1,3}$ and Jean-Philippe Bouchaud$^{1,4}$}

\date{\it\small
$^1$ Science \& Finance, 109--111, rue Victor Hugo,\\ 92535 Levallois {\sc
cedex},
France\\
$^2$ Institut de Physique Th\'eorique, \'Ecole Polytechnique F\'ed\'erale,\\
CH-1015 Lausanne, Switzerland\\
$^3$ Capital Futures Management, 109--111, rue Victor Hugo,\\ 92535 Levallois
{\sc cedex}, France\\
$^4$ Service de Physique de l'\'Etat Condens\'e,
 Centre d'\'etudes de Saclay, \\ Orme des Merisiers, 
91191 Gif-sur-Yvette {\sc cedex}, France \\}

%\date{\today}
\maketitle \abstract{We critically review recent claims that financial
crashes can be predicted using the idea of log-periodic oscillations
or by other methods inspired by the physics of critical phenomena.  In
particular, the October 1997 `correction' does not appear to be the
accumulation point of a geometric series of local minima.}

\vskip 2 true cm

It is rather tempting to see financial crashes as the analogue of
critical points in statistical mechanics, where the response to a
small external perturbation becomes infinite, because all the subparts
of the system respond cooperatively (a large proportion of the actors
in a market decide simultaneously to sell their stocks). If one
furthermore assumes that `log-periodic' corrections (for which there
is a recent upsurge of interest \cite{SReview}) are present, then one
can try to use the oscillations seen on markets as precursors to
predict the crash time $t_c$, which is the point where these
oscillations accumulate. Intriguing hints supporting this scenario
have been reported in \cite{S,F}, and more recently in \cite{F2,A2},
where it was even claimed that the October 1997 correction could be
predicted (see also \cite{Y}). As a proof of this, the implementation
of a winning strategy was reported in \cite{S2}. In view of the
considerable echo that these claims have enjoyed, in particular in the
physics community \cite{Europhys}, we feel that it is important to
temper the growing enthusiasm by discussing a few facts.

In general, the unveiling of a new phenomenon either results from a
strong theoretical argument suggesting its existence, or from
compelling experimental evidence. In the present case, there is no
convincing theoretical model which substantiates the idea that crashes
are critical points -- not even speaking about log-periodic
oscillations. On the `experimental' side, there has been only very few
crashes where this scenario (or any theoretical model for that matter)
can be tested. Hence, although suggestive, the empirical findings are
obviously not statistically significant.  The fact of correctly
predicting one event {\it ex ante} (the October 1997 correction
\cite{S2}) is clearly not enough to prove the theory right: many
`chartists' make a living by `recognizing' patterns on past charts of
prices and are on average right 50\% of the time. We want to publicly
disclose here the fact that on the basis of a log-periodic analysis, a
crash on the JGB (Japanese Government Bonds) for the end of May, 1995
was predicted \cite{DSunpub}. On this basis, one of us (JPA) bought
for \$1,000,000 of put options in early May 1995, a sum representing
less than 1\% of the amount under management of CFM (a fund management
company trading on the basis of statistical models). The crash did not
occur, and only a delicate trading back allowed to avoid losses. The
point is {\it not}\/ that the out-of-sample prediction failed (the
risk was deliberately taken), but rather that it worked one time and
failed the other, a fact that is not mentioned in \cite{S2}.
Obviously, this {\it does not}\/ mean that the method works 50\% of
the time (which would already be quite interesting); an apparent
success out of a least two trials is just not statistically
significant!

\begin{figure}
\centerline{\hbox{
\epsfig{figure=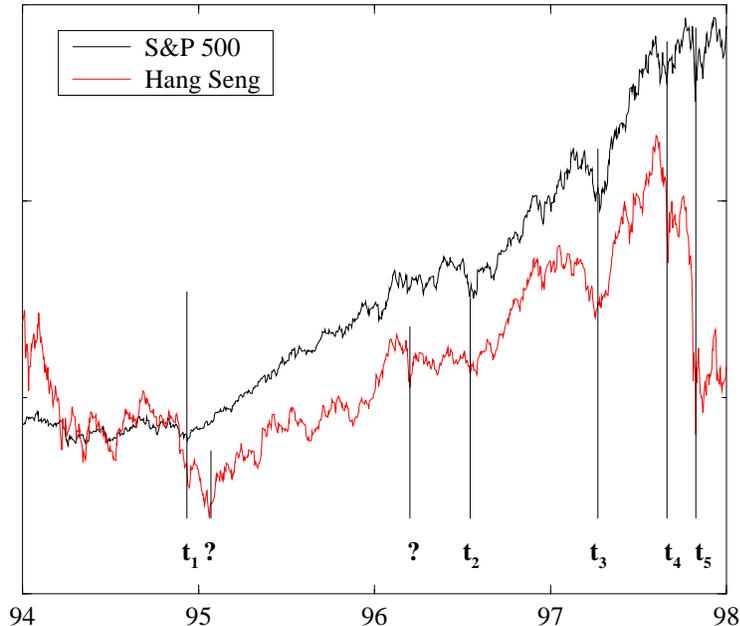,width=10cm}
}}
\caption{\small The S\&P 500 and the Hang Seng Index rescaled to fit on the
same (linear-log) graph, in the period 1994--1997.  The five local minima
$t_i$ are indicated
along with two `extra' minima for the Hang Seng.}
\end{figure}
The methodological procedure used to extract the `crash time' $t_c$ is
actually rather dubious, since a seven (or even nine) parameters fit
to noisy data is required, which is of some concern. Sometimes, the
time period over which the fit is performed is rather long, which
implies that a small dip occuring more than five years before the
crash must be seen as causally related to the crash, which is somehow
hard to believe. A more robust prediction of the log-periodic
scenario is that the price chart should exhibit a sequence of minima
(and maxima) at times $t_n$ such that:

\be
\frac{t_{n+1}-t_{n}}{t_{n}-t_{n-1}}= \lambda
\frac{t_{n}-t_{n-1}}{t_{n-1}-t_{n-2}} \qquad \lambda < 1
\ee

i.e.\ that the time lags between minima follows a geometric
contraction. The crash time is defined as the accumulation of these
minima. Looking at the S\&P chart between 1994 and the end of 1997
(see figure 1), one can identify five `major' minima, most of them
being identified as such by the press at the time, the last (at time
$t_5$) one being the October 1997 `crash' \cite{crash?}. (Of course,
there are also many local minima of less importance). One finds the
following time lags (in days): $t_2-t_1=403$, $t_3-t_2=182$,
$t_4-t_3=97$ and $t_5-t_4=44$. On the basis of the four first events,
one obtains three time lags and thus two estimates of $\lambda$,
namely $0.45$ and $0.53$, from which one estimates $t_5-t_4 = 48 \pm
4$: this is the `crash' prediction, which indeed worked well since the
observed $t_5-t_4=44$ \cite{us}. {\it But}, the scenario also predicts
that another drawdown should have occurred at $t_6$ such that $t_6-t_5
\simeq 20$, i.e, at the end of November 1997, and maybe another one
$9$ days later -- none of which occurred. One can of course argue that
the accumulation of crash times will be smeared out when one reaches
the day (or the week) time scale, but in any case a 6th minimum should
have been observed, otherwise the very notion of `critical point' is
empty. In view of the fact that three successive ratios $\lambda$ were
close to one another (and close to their $1987$ value \cite{87-97}),
the possibility of a crash occuring at the end of November was real;
conversely, its non existence throws serious doubts on the validity of
a log-periodic scenario. Furthermore, one can study the Hong-Kong
index which did experience a serious crash at the end of November
1997.  One finds that the above $t_n$, $n=1,5$ indeed corresponds to
rather deep minima; however, there are also at least two extra
`obvious' minimum between $t_1$ and $t_2$ which ruins the idea of a
constant $\lambda$ (see figure 1).

Using somewhat related ideas, another prediction of the October 1997
correction was reported in \cite{Y}. These authors also predicted a
13\% fall of the S\&P during March 1998 \cite{Y2}, which did not occur
(the index rose by 5\% instead). We have systematically tested their
method, where the price at the end of the next month is predicted on
the basis of the previous three to five months.  We studied the S\&P
from 1990 to the beginning of 1998, and found that, for example, the
prediction error using the 3 point procedure described in \cite{Y} is
about $10\%$ (see table 1), which is three times larger than the
simplest `no-change' prediction, i.e.\ that next month price is equal
to this month price!

%***Table with the results***
\begin{table}
\begin{center}
\begin{tabular}{||l|c|c||} \hline\hline
%\multicolumn{3}{||c||}{{\bf Monthly Strategies }} \\ \hline
Method:\ \hspace{1cm} \   & \hspace{1cm} $\overline{|r|}$ \hspace{1cm} &
\hspace{1cm}  $\sqrt{\overline{r^2}}$ \hspace{1cm} \\ \hline
no-change &  2.8  \% &   3.4 \% \\ \hline
3 months  &  7.5  \% &  10.0 \% \\ \hline
4 months  &  5.8  \% &  11.3 \% \\ \hline
5 months  &  12.3 \% &  17.0 \% \\ \hline
\end{tabular}
\end{center}
\caption[]{\small Results of the monthly trading strategies based on
the time series forecasting presented in \cite{Y,Y2}. The procedure
has been tested using 3, 4 and 5 past data points. The prediction
errors measured through $\overline{|r|}$ or $\sqrt{\overline{r^2}}$
(where $r$ is the relative error) are systematically at least three
times larger than the no-change prediction (i.e.\ constant price).
}
\end{table}

To answer the question raised in the title, we have argued that the
recent claims on the predictability of crashes are at this point not
trustworthy. This however does not mean that crash precursors do not
exist, an example could be a systematic increase of the volatility
prior to the crash. This general subject certainly merits further
investigations. Finance is a fascinating field with huge amounts of
money at stake. There is a danger that this might sometimes lead
physicists astray from minimal scientific rigor.

\vskip 1cm

We thank P. Cizeau and I. Kogan for discussions. We also thank
M. Ausloos, D. Sornette, D. Stauffer, N. Vandewalle and V. Yukalov for
comments on an earlier version of the paper.

\end{document}